\documentclass{ws-ijmpa}

\begin{document}

\markboth{Authors' Names}
{Instructions for Typing Manuscripts (Paper's Title)}

%
\catchline{}{}{}{}{}
%

\title{A NEW SEARCH STRATEGY FOR THE HIGGS BOSON
}

\author{JACQUES SOFFER}

\address{Centre de Physique Th\'eorique,
UMR 6207 \footnote{\uppercase{UMR} 6207 is \uppercase{U}nit\'e \uppercase{M}ixte
de \uppercase{R}echerche du \uppercase{CNRS} and of \uppercase{U}niversit\'es
\uppercase{A}ix-\uppercase{M}arseille \uppercase{I}
and \uppercase{A}ix-\uppercase{M}arseille \uppercase{II}, and of 
\uppercase{U}niversit\'e 
du \uppercase{S}ud \uppercase{T}oulon-\uppercase{V}ar, \uppercase{L}aboratoire
affili\'e \`a la \uppercase{FRUMAM}}\\
CNRS-Luminy Case 907, 13288 Marseille Cedex 9, France \\ 
E-mail: soffer@cpt.univ-mrs.fr}


\maketitle


\begin{abstract}
We propose a novel mechanism for
exclusive diffractive Higgs production $pp \to Hpp$, in which the Higgs carries a
significant fraction of the projectile proton's momentum. This mechanism will
then provide a clear experimental signal for Higgs production, due
to the small background in this kinematic region. The key
assumption underlying our analysis is the presence of intrinsic
charm (IC) and intrinsic bottom (IB) fluctuations in the proton
bound state, whose existence, at high light-cone momentum fraction $x$, has a
substantial and growing experimental and theoretical support.

\end{abstract}

\section{Introduction}

A central goal of the Large Hadron Collider (LHC) being built at
CERN is the discovery of the Higgs boson,  a key component of the
Standard Model, and whose discovery would also constitute the
first observation of an elementary scalar field.  A number of
theoretical analyses suggest the existence of a light Higgs boson
with a mass $M_H \lesssim 130~\mbox{GeV}$. Perhaps the most novel
production process for the Higgs is the exclusive
exclusive diffractive reaction, $pp \to p + H +
p$~\cite{2002hk}, where the + sign stands for a large
rapidity gap (LRG) between the produced particles.  If both
protons are detected, the mass and momentum distribution of the
Higgs can be determined. The TOTEM detector~\cite{2004gt}
proposed for the LHC will have the capability to detect exclusive
diffractive channels.

The detection of the Higgs via the exclusive diffractive process $pp
\to p + H + p$, has the advantage that it  does not depend on a
specific decay mechanism for the Higgs. The branching ratios for
the decay modes of the Higgs can then be individually determined
by combining the measurement of $\sigma(pp \to p + H + p)$ with
the rate for a specific exclusive diffractive final states
$B_f~\sigma(pp \to p + H_{\to f} + p)$. This is in contrast to the
standard inclusive  measurement  where one can only determine the
product of the cross section and  branching ratios $B_f~ \sigma(
pp \to H_{\to f} X).$

The existing theoretical estimates for exclusive diffractive Higgs
production are based on the gluon-gluon fusion subprocess, where
two hard gluons couple to the Higgs $(gg \to
H)$~\cite{2002hk}. A third gluon is also exchanged  in
order that both projectiles remain color singlets. Perturbative
QCD then predicts $\sigma(pp \to p + H + p) \simeq 3~\mbox{fb}$
for the production of a Higgs boson of mass 120 GeV at  LHC
energies, with a factor of 2 uncertainty \cite{2002hk}.
Since the annihilating gluons each carry a small fraction of the
momentum of the proton, the Higgs is primarily produced in the
central rapidity region.

In a recent work \cite{bkss}, we proposed a novel mechanism for exclusive diffractive
Higgs production in which the Higgs is produced with a significant
fraction of the projectile's  momentum.  The key assumption underlying
our analysis is the presence of intrinsic charm (IC) and intrinsic
bottom (IB) fluctuations in the proton bound state~\cite{1981se,1980pb},
whose existence at high $x$  has a substantial and growing experimental and
theoretical support.  The virtual Fock state $|uud Q \bar Q>$ of one of
the incident protons has a long lifetime at high energies and
can be materialized in the collision by the
exchange of  gluons. The heavy quark and antiquark from the same
projectile then coalesce  to produce the Higgs boson at large $x_F.$

It was originally suggested in Ref.~[4,5] that there is a $\sim
1\%$ probability of IC Fock states in the nucleon; more recently,
the operator product expansion has been used to show that the
probability for Fock states in light hadron to have an extra heavy
quark pair of mass $M_Q$ decreases only as $\Lambda^2_{QCD}/M_Q^2$
in a non-Abelian gauge theory \cite{2000ee}.  In the case
of  Abelian QED, the probability of an intrinsic heavy lepton pair
in a light-atom such as positronium is suppressed by
$\mu^4_{bohr}/M_\ell^4$ where $\mu_{bohr}$ is the Bohr momentum.
The quartic QED scaling corresponds to the dimension-$8$
Euler-Heisenberg effective Hamiltonian  $F^4/ M^4_\ell$ for
light-by light scattering mediated by heavy leptons.  Here $F_{\mu
\nu}$ is the electromagnetic field strength. In contrast, the
corresponding effective Hamiltonian in QCD $G^3/M^2_Q$ has
dimension $6.$ The maximal probability for an intrinsic heavy quark Fock state occurs
for minimal off-shellness;  i.e., at minimum invariant mass squared
${\cal M}^2 = \sum^n_{i=3} (m^2_i +{\vec  k}^2_{\perp i})/ x_i  .$ Thus
the dominant Fock state configuration is $x_i \propto m_{\perp i}$
where $m^2_{i \perp} =m^2_i + {\vec k}^2_{\perp i};$ i.e.,  at equal
rapidity.   Since all of the quarks tend to travel coherently at  same
rapidity in the $|uudQ\bar Q>$ intrinsic heavy quark Fock state,
the heaviest constituents carry the most momentum~\cite{1981se,1980pb}.

\section{Experimental Evidence for Intrinsic Charm}

The most direct test of intrinsic charm is the measurement of the
charm quark distribution $c(x,Q^2)$ in deep inelastic scattering
$\ell p \to \ell^\prime c X.$  The only experiment which has
looked for a charm signal at large $x_{bj}$ is the European Muon
Collaboration (EMC) experiment~\cite{1982tt} which used
prompt muon decay in deep inelastic muon-proton scattering to tag
the produced charm quark. The EMC data show a distinct excess of
events in the charm quark distribution at $x_{bj}  >  0.3,$ at a
rate at least an order of magnitude beyond lowest  predictions
based on gluon splitting and DGLAP evolution. More recent
next-to-leading order (NLO) analyses~\cite{1995jx} show
that  an intrinsic charm component, with probability of order
$1\%$,  is needed to fit the EMC data in the large $x_{bj}$
region.

The existence of the rare double IC Fock state such as
$|uudc\bar{c}c\bar{c}>$ leads to the production of two
$J/\psi$'s~\cite{1995tf} or a double-charm baryon state at
large $x_F$ and small $p_T$.   Double $J/\psi$ events at a high
combined $x_F \ge 0.8$ were in fact observed by
NA3~\cite{1982ae}. The observation of the doubly-charmed
baryon $\Xi^{+}_{cc}(3520)$ has been confirmed recently by SELEX
at FNAL~\cite{2004hi};  the presence of two charm
quarks at large $x_F$ has, indeed, a natural IC interpretation.
\footnote{Additional relevant experimental facts are listed in Ref.~[3]}

\begin{figure}
\centerline{\psfig{file=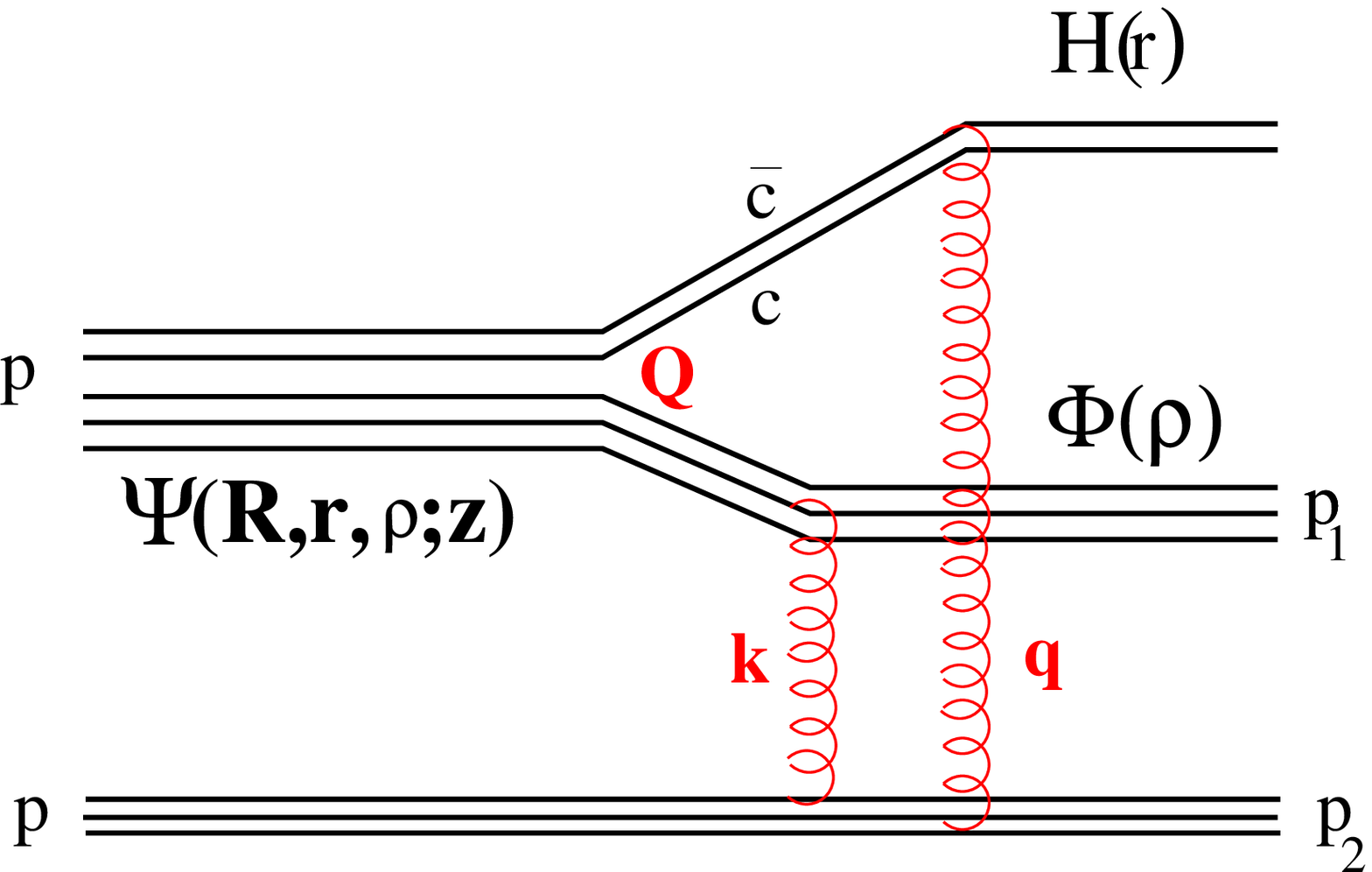,width=8cm}}
\vspace*{8pt}
\caption{The exclusive diffractive production of the Higgs boson.}
\end{figure}

\section{Higgs Production}
 We will now explain, using Fig.~1, how the exclusive 
diffractive production arises with the
required color structure in the final state.  As noted above, the
IC Fock state of the projectile (upper) proton has a $1\%$
probability to fluctuate to the state with the color structure
$|[uud]_{8_C}[\bar c c]_{8_C}>$. It has a long coherence length in
a high energy collision $\propto {s/{\cal M}^2 M_p}$, where
$\cal{M}$ is the total invariant final mass, which is much larger
than the initial mass. In a $pp$ collision, two soft gluons must
be exchanged in order to keep the target (lower) proton intact and
to create a rapidity gap. The two gluons will couple from the
target to the large color dipole moment of the projectile IC Fock
state. For example, as shown in Fig.~1, one gluon can be
attached to the $d$ valence quark spectator in $|[uud]_{8_C}[\bar
c c]_{8_C}>$, changing its color, and the other one can be
attached to the $\bar c $, changing also its color. The net effect
of this color rearrangement is the same as a single gluon exchange
between the two color-octet clusters. The $\bar c c$ and the $uud$
thus can emerge as color singlet because of the gluonic exchange.
The $[\bar c c]_{1_C}$ now can couple to the $J/\psi$ color
singlet, or to a $Z$ or to a $H$. Meanwhile the color-singlet
$uud$ gives rise to the scattered proton, and we have the two
required rapidity gaps in the final state. Notice that the $x_F$
distribution of the produced particle is approximately the same as
the distribution of the $[\bar c c]$ inside the proton.
Unfortunately the coupling of the gluon to all the different
quarks brings in a form factor which vanishes at zero momentum
transfer, and which therefore gives an important suppression
factor.

The cross section of exclusive diffractive production of the
Higgs, $pp\to ppH$, can be estimate in two-gluon approximation in
accordance with the Feynman graph shown in Fig.~1.
 Here we assume the presence in the proton of an intrinsic charm (IC)
component, a $\bar cc$ pair which is predominantly in a
color-octet state, and which has a nonperturbative origin and
mixes in the proton with the $3q$ valence quark component.
Correspondingly, like in charmonium, the mean $\bar cc$ separation
should be much larger than the size $1/m_c$ of perturbative $\bar
cc$ fluctuations. In Fig.~1, $\Psi$ denotes the light-cone wave function
of the IC component of the initial proton, properly normalized, $H$ is the wave
function of the color singlet $[\bar c c]_{1_C}$ in the produced Higgs and $\Phi$
stands for the wave function of the final proton $p_1$.
The details and different
aspects (energy dependence, absortive corrections, etc...) of the calculation
will be reported in Ref.~[3], but we anticipate
that the final result will turn out to be rather small. 
Various possibilities to get a larger cross section will be also discussed,
in particular, heavy flavors contribution or nuclear enhancement.

\section*{Acknowledgments}

I am glad to thank S.~J.~Brodsky, B.~Kopeliovich and I.~Schmidt, my collaborators
on this work. I am also grateful to the organizers of this exciting International
Conference on QCD and Hadronic Physics,  for the invitation
and for the opportunity to give this talk.

\end{document}